# Multispectral CCD-in-CMOS Time Delay Integration imager for high resolution Earth observation

**Swaraj Bandhu Mahato[1], Steven Thijs[1], Jonas Bentell[1], Linkun Wu[1], Klaas Tack[1], Pierre Boulenc[1], Dorian Lasnet[2], Renaud Van Langendonck[2] and Piet De Moor[1]**

[1]imec, Kapeldreef 75, 3001 Leuven, Belgium
[2]Aerospacelab, Belgium
Phone: +3216283628 Email: swaraj.mahato@imec.be

**Abstract:** Many future small satellite missions are aimed to provide low-cost remote sensing data at unprecedented revisit rates, with a ground resolution of less than one meter. This requires high resolution, fast and sensitive line-scan imagers operating at low power consumption and ideally featuring spectral sensitivity. In this paper we present comprehensive characterization results of our 7 band Back-Side Illuminated (BSI) CCD-in-CMOS sensor with a pixel pitch of 5.4 µm. We have extensively characterized the key performance parameters of our CCD-in-CMOS sensor, such as quantum efficiency (QE), full well capacity (FWC), read noise, conversion gain, non-linearity, dark current etc. Novelty of this device is the combination of 7 TDI bands on the same imager allowing simultaneous multispectral TDI capture. Glass-based broadband filters with a typical bandpass width of about 100 nm have been developed and glued together to form a filter assembly of 6 bandpass filters and one panchromatic channel. Multispectral capability of this sensor is particularly interesting for Low Earth Observation (LEO) applications such as environmental monitoring, precision agriculture, disaster detection and monitoring. To highlight its advantages for use in vegetation observation, we demonstrated a fake leaf and a real leaf imaging using a 7 band BSI sensor with integrated filters operating in 7-band mode at 15 kHz.

## 1. INTRODUCTION

Recent technological advancement of space component miniaturization has enabled small spacecrafts which can assure a fast and affordable access to space for Low Earth Observation. Although traditionally Time-Delay-Integration (TDI) CCDs have been the only popular choice of image sensor for Earth observation, they have some significant drawbacks for small satellite use. They consume a lot of power and a staggered multi-chip solution is required when spectral information is desired, leading to alignment inaccuracies and complex bulky optics. Imec first introduced its monolithic CCD-in-CMOS technology at IEDM 2014 [1]. It combines the benefits of a classical CCD TDI with the advantages of CMOS System-On-a-Chip (SoC) design. Imec's CCD-in-CMOS technology is continuously being tuned to reach high pixel performance. Our current 7 band Back-Side Illuminated (BSI) CCD-in-CMOS sensor has 256 TDI stages per band, with 4096 columns at a pixel pitch of 5.4 µm. Small pixel size allows a high swath width and high resolution without compromising on the satellite speed. Novelty of this device is the combination of 7 TDI bands on the same imager allowing simultaneous multispectral TDI capture. Glass-based broadband filters with a typical bandpass width of about 100 nm have been developed and glued together to form a filter assembly of 6 bandpass filters and one panchromatic channel (ESA AO/1 9315/18/NL/AR MICROMHIDE project). Multispectral capability of this sensor is particularly interesting for Low Earth Observation (LEO) applications such as environmental monitoring, precision agriculture, disaster detection and monitoring. A detailed and accurate knowledge of the performance parameters (e.g. quantum efficiency (QE), full well capacity (FWC), read noise (RN), conversion gain (CG), non-linearity, dark current (DC) etc.) is of great interest for space imaging. Many applications demand accurate knowledge of the quantitative image data, which highly depend on the performance of the image sensor.

In this paper, a comprehensive characterization of our CCD-in-CMOS TDI image sensor is presented. We extensively evaluated the key performance parameters of the TDI sensor in terms of its capability in LEO applications for small satellite missions.

## 2. TDI SYSTEM-ON-CHIP

As reported in [3], Imec has developed a Digital-Out CCD-in-CMOS TDI imager. Our TDI imager features a specifically developed technology [1-2] which combines the benefits of a classical CCD TDI with the advantages of CMOS System-On-a-Chip (SoC) design. This CCD-in-CMOS platform was realized by adding a few process modules to a standard 0.13 µm CMOS process flow containing dual gate oxide nMOS and pMOS transistors (1.2 and 3.3V) enabling monolithic integration of CCD row drivers and fast 12-bit analog-to-digital converters (ADCs) at each column [4]. The pixels are described in detail in [2]. We used dedicated buried channel, isolation, well and junction implants for the CCD elements and readout transistors to ensure CMOS compatible operating voltages. To limit the dark current, no Shallow-Trench-Isolation (STI) is used for channel separation. The block diagram of the sensor is shown in Figure 1. As each band uses individual on-chip sequencers and CCD drivers for the four-phase CCD pixels, the 7 bands can be configured independently. This gives an individual TDI stage selection capability per band (with single stage resolution) to match the scene irradiance and to center the dynamic range to the relative specific spectral intensity of the scene and the transmission losses of filters. Bi-directional readout is possible thanks to top and bottom dedicated charge to voltage conversion stages.

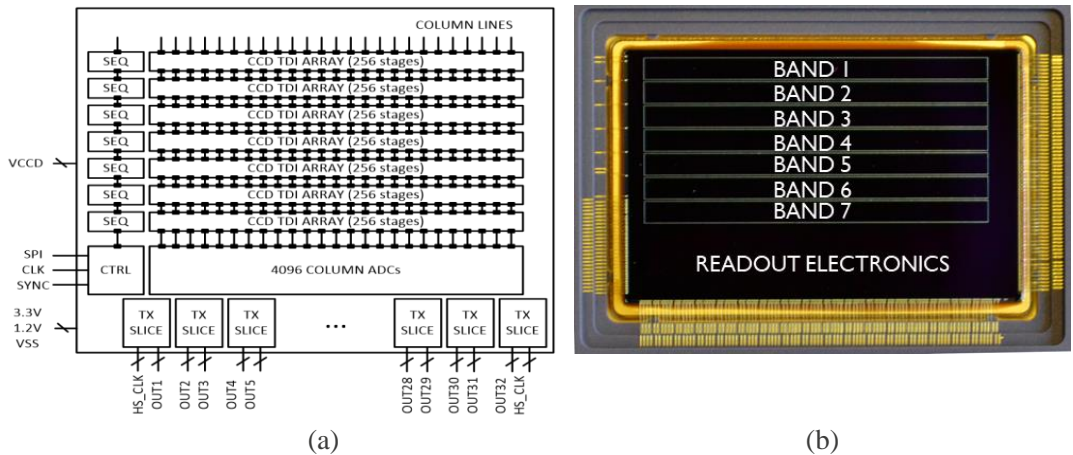

*Figure 1: 7-band CCD-in-CMOS TDI block diagram (a) and a Back-Side-Illuminated sample (b).*

## 3. PERFORMANCE CHARACTERIZATION

### 3.1 Dark Signal Transfer Curve (DSTC)

Dark signal transfer curve has been measured at room temperature by means of flush/integration/readout sequences with varying integration time in the dark. Figure 2 depicts the DSTC, where noise is plotted as a function of the mean signal, used to determine

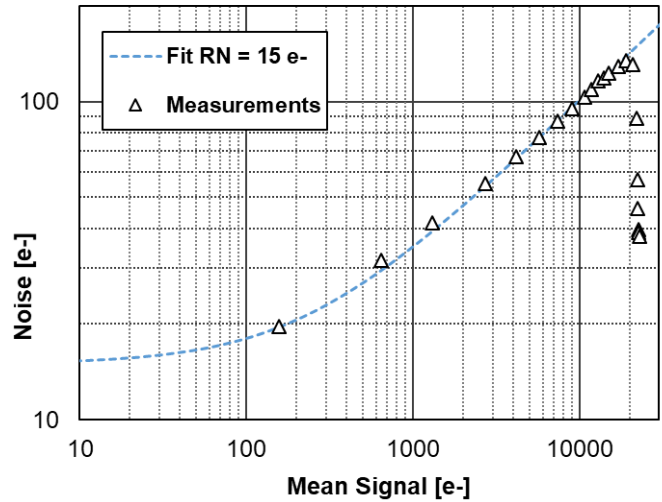

*Figure 2: Dark signal transfer curve (DSTC) of the TDI sensor where noise is plotted as a function of the mean signal. Dark Signal Transfer Curve measured at 115 kHz line rate on a single CCD band.*

Conversion Gain (CG), Full Well Capacity (FWC) and noise floor. The conversion gain is about 1.97 DN/$e^-$ and the saturation FWC is about 23000 $e^-$. A noise floor of about 15$e^-$ dominated by the read noise is measured. Maximum noise is reached for a signal level of 20000 $e^-$. This is referred as a blooming FWC when electrons get redistributed in the neighboring pixels within the same column [5]. The Dynamic Range (DR) of the sensor is calculated from the measured read noise floor and the FWC to be about 62 dB.

### 3.2 Signal Response and Linearity

The term linearity implies that the response should have a linear relationship to varying brightness and integration times. The transfer function between the incident photon-signal and the final digitized output should be linear. The linearity is measured by exposing the image sensor to a constant flat illumination source for different exposure times up to the saturation level.

The image sensor response curve and nonlinearity are depicted in Figure 3, respectively. The nonlinearity of an image sensor indicates the deviation of the output signal from an ideal straight fit of the response curve (linear regression of the response). In our measurement, the linear fit was done between 5 % and 95 % of the saturation level which is shown in the Figure 3(a). The apparent nonlinearity for each integration time is measured and normalized by the maximum mean signal level, which is called as a relative non-linearity (Figure 3(b)) of the sensor response. The relative non-linearity of this sensor is within 1% for the signal levels ranging from 5 % to 95 % of the saturation level.

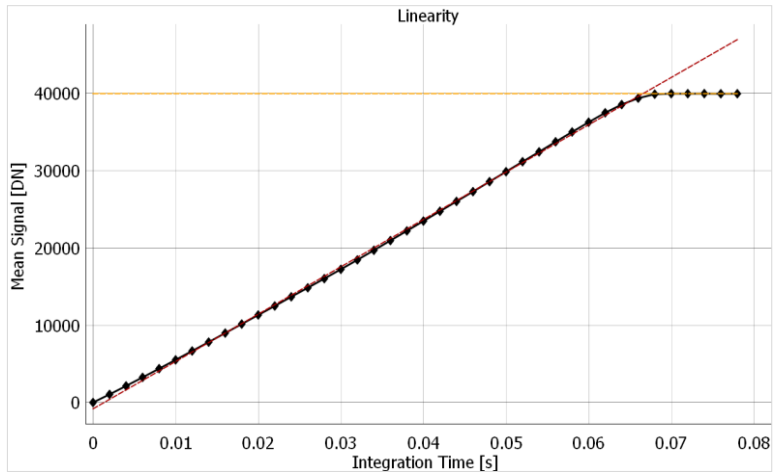

(a)

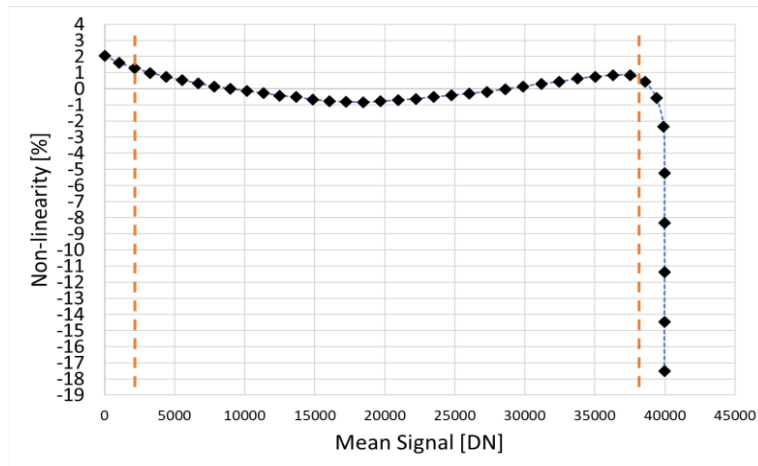

(b)

*Figure 3: (a) Response curve of the TDI sensor. The linear fit is done between 5 % and 95 % of the saturation level. (b) Relative non-linearity of the TDI sensor as a function of the mean signal. The relative non-linearity is within 1% for 5 % to 95 % of the saturation level indicated by the orange vertical lines.*

### 3.3 Dark Current

For low-light applications, any noise limits the sensitivity of the image sensor. The dark current is an electric leakage current generated in the photodiode even when the detector is not exposed to any light. The generation of dark electrons is a thermally activated process and therefore strongly varies with temperature. We measure the dark current as a function of temperature for each pixel by fitting a linear slope to the increasing signal value with an exposure sequence. The measurements are repeated for various temperature settings of the thermal chamber. Figure 4 plots the dark current versus the operating temperature. The measured dark current is referred to the photocurrent using conversion gain of 1.97 DN/e-. The calculated dark current doubling temperature is 8.3°C. The dark current at room temperature is 5.6 nA/cm² (10200 e-/s/pix).

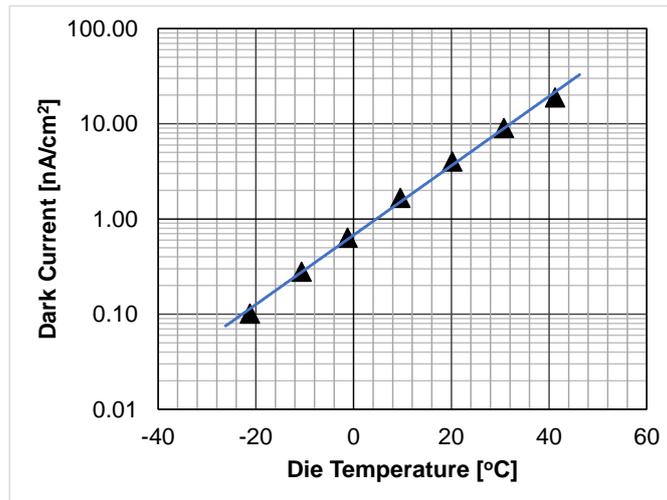

*Figure 4: Dark current versus die temperature. Calculated dark current doubling temperature is 8.3ºC.*

### 3.4 Charge Transfer Efficiency (CTE)

Charge Transfer Inefficiency (CTI = 1 - CTE) measurements have been carried out on the pixel test vehicle featuring the same pixel design as dedicated test structures are required for this type of measurement. Characterization has been performed for line rates ranging from 2 kHz up to 1 MHz at various temperatures and signal levels up till linear Full Well Capacity.

In Figure 5, CTI degradation because of traps is visible at low line rate for signal levels below 8000 electrons [5-6]. A slow transfer allows traps to capture electrons before the next transfer takes place and release them in subsequent transfers. CTI is degraded above saturation FWC and changes with the line rate. The faster the CCD is clocked, the higher is the signal that can be transported with negligible losses. A CTI lower than $1e^{-5}$ (CTE > 99.999%) is measured over the full line rate and temperature range.

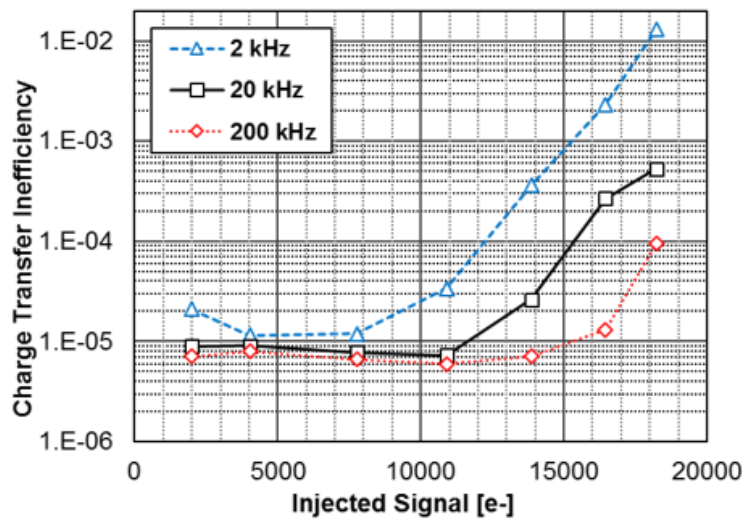

*Figure 5: CTI vs injected signal level for various line rates at 25°C.*

### 3.5 Quantum Efficiency

Our backside processing and Anti-Reflective Coating (ARC) have been optimized to enhance Quantum Efficiency (QE) for the Red-Green-Blue (RGB) spectrum with 96 % peak QE at 510 nm wavelength and for the ultra-violet spectrum with 93 % peak QE at 310 nm wavelength on a 5.8 μm thick substrate, as shown in Figure 6.

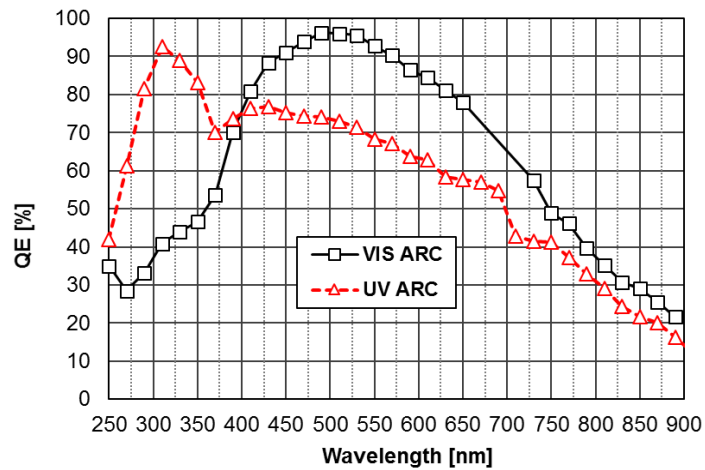

*Figure 6: Quantum Efficiency vs. wavelength for BSI TDI sensor with 5.8 μm thick substrate and different Anti-Reflective Coatings (ARC).*

### 3.7 Radiation test

Under the framework of the MICROM-HIDE ESA project and in collaboration with Aerospacelab, a preliminary radiation campaign was performed. The test objective was to do a first sanity check of the sensor under radiation, paving the way for a more extensive characterization campaign. Figure 7 shows the radiation test setup at the cyclotron facility in Louvain-La-Neuve, Belgium. In the proton test higher energy protons will mostly go through the active material without doing any damage. Hence, there was no point in using the highest available voltage, but instead we used a retarded beam and a final energy of ~10 MeV. Fluence was selected to ~5E10 /cm$^2$ with a flux that is approximately four orders of magnitude lower, 1.4E7. There was no latch-up (hard or micro), or Single Event Functional Interrupt (SEFI) observed under proton test. In heavy ion test, Linear energy transfer (LET) was varied from 0 to ~45 MeV.cm$^2$.mg by selecting different species from the cocktail of the cyclotron. For each species and angle, we counted the latch-up events. Our TDI Sensor passed a first batch of radiation tests from a LET of 1.3MeV to 45.8MeV and was still operational afterward. Micro latch-ups are the main issue. They appear at all the energy levels tested and reduce the availability of the chip. Further tests need to be done to characterize the sensor performance under radiation.

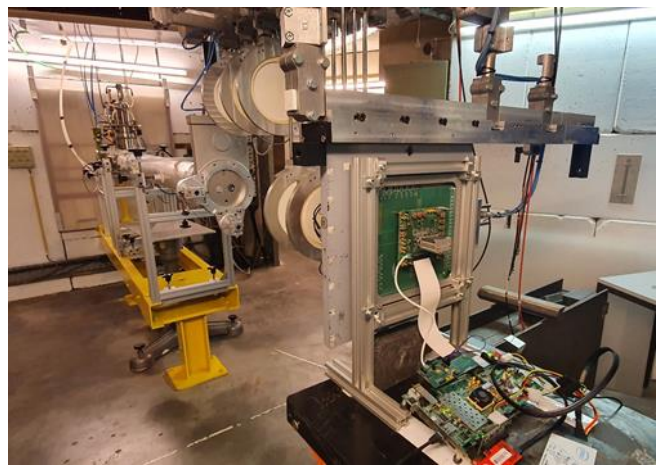

*Figure 7: Radiation test setup at the cyclotron facility in Louvain-La-Neuve, Belgium.*

### 3.6 Multi-spectral TDI imaging

To demonstrate the multi-spectral TDI imaging, we realized our 7-band CCD-in-CMOS TDI full System-on-Chip imager with 7 glass filters attached on top of the imager at die level with dimensions matched to the available TDI bands on the chip [7]. Each of these broadband filter covers approximately 100 nm spectral width. They have been glued together with black epoxy to shield the chip area outside the CCD bands. It forms a filter assembly of 6 bandpass filters and one panchromatic channel (depicted in Figure 8) and integrated with controlled alignment on a demo sensor.

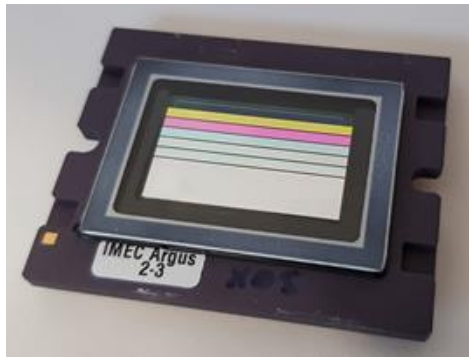

*Figure 8: 7-band assembled glass filters, direct integrated onto the multi-band CCD-in-CMOS TDI imager.*

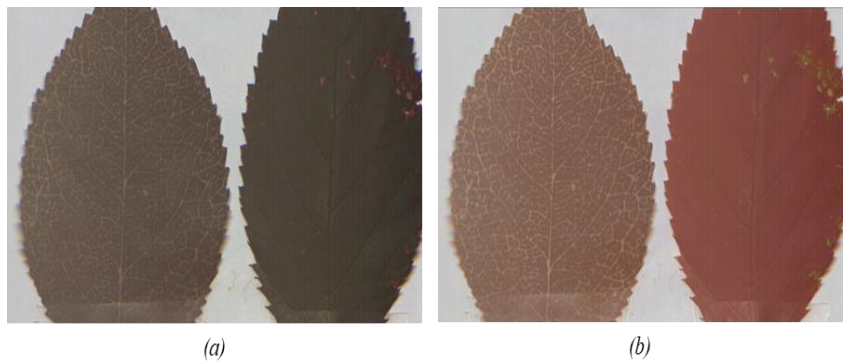

*(a)* *(b)*

*Figure 9:(a) The fake leaf (left) and the real leaf (right) can hardly be distinguished by RGB only. (b) The real leaf (right) has high reflectance above the chlorophyll red edge and looks bright red.*

In Figure 9, images were acquired using a 7 band BSI sensor with integrated filters operating in 7-band mode at 15 kHz line rate to highlight its advantages for use in vegetation observation. The RGB channels were used to construct an image Figure 9(a) that contains both the fake leaf and the real leaf side-by-side while another similar image Figure 9(b) is constructed by composing 2 red images on either side of the vegetation red edge and the green channel. The fake leaf and the real leaf can hardly be distinguished by RGB only while in the second image the real leaf has high reflectance above the chlorophyll red edge and looks bright red. Optimized stage selection was applied to each individual band to maximize SNR while avoiding saturation.

### 3.7 Conclusions

In this paper we presented comprehensive characterization results of the 7-band CCD-in-CMOS TDI full System-on-Chip imager. We have extensively characterized the key performance parameters of our CCD-in-CMOS sensor, such as quantum efficiency (QE), full well capacity (FWC), read noise, conversion gain, non-linearity, dark current etc. Multispectral capability of this sensor has been demonstrated using a 7 band BSI sensor with integrated filters operating in 7-band mode at 15 kHz line rate. These results strengthen the state-of-the-art multispectral capabilities of Imec's 7-band CCD-in-CMOS TDI imager for high resolution Low Earth Observation (LEO) applications.

### 4. ACKNOWLEDGEMENTS

This project has received funding from the Electronic Component Systems for European Leadership Joint Undertaking under grant agreement No 662222. This Joint Undertaking receives support from the European Union's Horizon 2020 research and innovation program and Belgium, Netherlands, Greece, France.

The filter integration is part of the ESA AO/1 9315/18/NL/AR contract MICROM-HIDE: Technologies for Microsatellites Multispectral High-Definition Imager.